# Is Unitarity of CKM Matrix Violated?

Mohammad Saleem

Theory Group, Centre for High Energy Physics,
University of the Punjab, New Campus, Lahore.

**Abstract**

The unitarity of the CKM matrix is examined in the light of the latest available accurate data. The analysis shows that for the elements in the first row, there are two types of experimental measurements leading to different results for the violation of unitarity. A conclusive result cannot be derived at present. Only more precise data can determine whether the CKM matrix opens new vistas beyond the standard model or not. The experimental data for the elements in the second row are not as precise as for those in the first row, and within the measured uncertainty, the unitarity is not violated. The desired data is not available for the elements in the third row.

The quark eigenstates of weak interactions do not correspond to the quark mass eigenstates. These are related to each other through a 3 x 3 unitary matrix, say V, generally known as CKM matrix. Since the matrix V is unitary, the sum of the squares of the magnitudes of the matrix elements for each row and column must be unity. This can be checked by taking elements in various rows and /columns and adding the squares of their magnitudes. Thus each of the six equations will verify whether the unitarity is violated or not. This violation of unitarity, if established, may lead to the conclusion that the standard model for three generations of fermions is not valid and it will then open new vistas.

The precise measurements are yet confined to the elements of the first row for which the unitarity requires

$$|V_{ud}|^2 + |V_{us}|^2 + |V_{ub}|^2 = 1. \qquad (1)$$

The value of $|V_{ud}|$ has been accurately measured by using superallowed $\beta$ decays where only the weak vector current contributes and by giving a very careful consideration to radiative corrections. A very accurate value obtained [1] by this technique is

$$|V_{ud}| = 0.9740 \pm 0.0005.$$

In fact, some of the nuclear collections are so difficult to measure that the Particle Data Group has doubled the error in $|V_{ud}|$. To avoid these difficult to measure nuclear corrections, Abele et al. [2] have recently written a very important paper in which they have given the result by using a new technique to measure $|V_{ud}|$. Instead of considering the superallowed β decay, they have used the spectrometer PEKROE II to derive $|V_{ud}|$ by measuring the neutron decay data. Therefore the technique for checking equation (1) is based only on particle physics phenomena where, of course, theoretical uncertainties are fewer and smaller. However, it depends upon $g_A/g_V$, the ratio of the axial and the vector coupling constants, as well as the neutron lifetime. As the experimental progress had been made using highly polarized cold neutron beams together with improved detectors, it was capable of competing with nuclear β decays in extracting a value for $|V_{ud}|$, whilst avoiding the problems linked to nuclear structure. This group found that

$$|V_{ud}| = 0.9713 \pm 0.0013.$$

The main experimental errors are due to spin polarization, background subtraction and detector response. By virtue of this result, the unitarity test of the CKM matrix can be performed by using only particle physics data.

To determine the modulus of the element $V_{us}$, the decays of $K^+$ and $K^0_L$ to $\pi^0 e^+ \nu_e$ and $\pi^- e^+ \nu_e$ were considered. Since the decays occur via pure vector currents, they involve only second order correction terms in SU(3) breaking symmetry. By using the chiral perturbation theory for corrections, the result obtained was [3]

$$|V_{us}| = 0.2196 \pm 0.0023.$$

It may be mentioned that the analysis of semi-leptonic hadronic decays for measuring $|V_{us}|$ involve larger theoretical uncertainties since these are governed by vector as well as axial vector currents. The SU(3) symmetry breaks in the first order and consequently the larger theoretical uncertainties. But it still gives a result, $|V_{us}| = 0.222 \pm 0.003$ [4], that may be considered as consistent with the above result. This is interesting to note that this measurement was made long time ago and in view of the progress in experimental techniques, it is expected that more accurate results would be obtained which may be extremely helpful in determining the unitarity of the CKM matrix.

$V_{ub}$ is one of the smallest and least well-known CKM matrix elements in its first row. The magnitude of this element can be obtained by using the exclusive b → ulν decays. To determine the magnitude of the element $V_{ub}$, the hadronic form factors have to be used and are obtained from theory. Very recently, the BABAR Collaboration [5] have used five different form factors for computations and the combined result obtained by them by collecting about 55 million B $\bar{B}$ pairs in a preliminary measurement of the CKM matrix element $|V_{ub}|$ is

$$|V_{ub}| = (3.69 \pm 0.23 \pm 0.27 \,^{+0.40}_{-0.59}) \times 10^{-3}.$$

The quoted errors are statistical, systematic and theoretical, respectively. In fact, different results have been obtained by using different form factors. The above result is the average value.

The measurements made by several groups are shown in Table 1 [5] and reflect the need for more accuracy- less uncertainty- in the measurement of the modulus of this matrix element.

Table 1

| Collaboration | $|V_{ub}| \times 10^3$ |
|---|---|
| BABAR (exclusive) | $3.69 \pm 0.23 \pm 0.27\,^{+0.40}_{-0.59}$ |
| CLEO (exclusive I) | $3.23 \pm 0.24\,^{+0.23}_{-0.26} \pm 0.58$ |
| CLEO (exclusive II) | $3.25 \pm 0.14\,^{+0.21}_{-0.29} \pm 0.55$ |
| CLEO (inclusive) | $4.08 \pm 0.63$ (stat + syst + theo) |
| LEP (inclusive) | $4.09 \pm 0.68$ (stat + syst + theo) |

Kim et al. [6] have also studied semi-inclusive charmless decays $B \to \pi X$, where X does not contain a charm quark or antiquark. They investigated the possibility of extracting the modulus of $V_{ub}$ from these processes. Of course, in general these decays are expected to have less hadronic uncertainty and larger branching ratios as compared to the exclusive decays. The mode $\bar{B}^0 \to \pi^- X$ turns out to be particularly useful for the determination of the modulus of the CKM matrix element $V_{ub}$. For the branching ratio $\beta = (1.0 \pm 0.1) \times 10^{-4}$, they expected

$$|V_{ub}| = (3.7 \pm 0.47) \times 10^{-3}.$$

This is in close agreement with the average value obtained by the BABAR Collaboration.

Bornheim et al. [7] have recently measured the magnitude of this element and with a number of uncertainties have found the value as

$$|V_{ub}| = (4.05 \pm 0.89) \times 10^{-3},$$

but this value is not precise.

If we consider only the data obtained from particle physics phenomena, we find

$$|V_{ud}|^2 + |V_{us}|^2 + |V_{ub}|^2 = 0.9917 \pm 0.0028.$$

The value differs from the unitarity constraint by $0.0083 \pm 0.0028$. This is about $3\sigma$ times away from the stated error. Hence, it conflicts the prediction of the standard model and if confirmed will give us a new perspective.

On the other hand, if we take the value of $|V_{ud}|$ as

$$|V_{ud}| = 0.974 \pm 0.0005,$$

it is found that

$$|V_{ud}|^2 + |V_{us}|^2 + |V_{ub}|^2 = 0.9969 \pm 0.0024.$$

This is only about 1.3 standard deviation from the unitarity and the result is therefore consistent with the standard model.

Therefore, as far as the first row of the CKM matrix is concerned, at present, nothing can be conclusively said about the violation of the unitarity of the CKM matrix. More precise measurements are required and are being eagerly awaited.

Let us next consider the second row of the CKM matrix. First we consider the element $V_{cd}$, the element in the second row and the first column. The neutrino and antineutrino production of charm off valence d quarks has been used to obtain $|V_{cd}|$. The average value of $\bar{B}_c |V_{cd}|^2$, where $\bar{B}_c$ is the semileptonic branching fraction of the charmed hadrons produced, obtained using the dimuon production of CDHS group and CCFR Tevatron experiments [8-9] is $(0.49 \pm 0.05) \times 10^{-2}$. Along with the data [10] on the mix of charmed particle species produced by neutrinos, and $\bar{B}_c = 0.099 \pm 0.012$ [9], this yields

$$|V_{cd}| = 0.224 \pm 0.016.$$

The value of the element related with the quarks of the second generation is not known with precision. In semileptonic decays of D mesons, there is the uncertainty of about $\pm 18\%$ on $|V_{cs}|$ and largely this comes from poorly known form factors. An alternative method of the measurement of this modulus stems from the hadronic decays of charged weak bosons $W^{\pm}$ produced in $e^+e^-$ interactions at the upgraded LEP2 collider. In the decays $W^{\pm} \rightarrow q_1 \bar{q}_2$, the coupling of the $W^{\pm}$ to $q_1$ and $\bar{q}_2$ is proportional to the appropriate CKM matrix element. Hence $|V_{cs}|$ can be determined from the measured rate of $W^{\pm} \rightarrow c\bar{s}$ decays. It may be mentioned that additional information can be obtained by tagging the flavours of the jets produced in hadronic $W^{\pm}$ decays [11].

DELPHI Collaboration has used the decays of $W^{\pm}$ bosons produced at LEP 2 to measure $|V_{cs}|$ [11]. The values for $|V_{cs}|$ were extracted from the measured hadronic branching ratio of $W^{\pm}$ decays and by tagging the flavour of hadronic jets produced in $W^{\pm}$ decays. Applying the two methods to the sample of approximately 100 W pairs collected during 1996 at energies of 161 and 172 GeV, DELPHI obtained

$$|V_{cs}| = 0.90 \pm 0.17 \text{ (stat)} \pm 0.04 \text{ (syst)}$$

and $\quad |V_{cs}| = 0.94 ^{+0.32}_{-0.26} \text{ (stat)} \pm 0.13 \text{ (syst)}.$

This surpasses the precision of the combination of all previous measurements [8, 12-14]. It may be emphasized that, instead of theoretical uncertainties, the uncertainty in the result is dominated by the statistical error.

In 1998, the ALEPH Collaboration also reported preliminary results of the measurement of $|V_{cs}|$ in hadronic W decays [15]. The inclusive charm production rate in W decays was measured from a study of kinematic properties of final state particles. Using 67.7 pb$^{-1}$ of data collected by ALEPH in 1996 and 1997 at center of mass energies 170 and 184 GeV and studying the two channels $W^+W^- \rightarrow 4q$ and $W^+W^- \rightarrow l\nu q\bar{q}'$, the value of $|V_{cs}|$ was found as

$$|V_{cs}| = 1.00 \pm 0.10 \text{ (stat)} \pm 0.06 \text{ (syst)}.$$

This significantly improved the then world average value of 1.01 ± 0.18 [4].

The final result was reported in September 1999 [16]. Using a charm tag based on the properties of jets produced in W decays, it was found that

$$|V_{cs}| = 1.00 \pm 0.11 \text{ (stat)} \pm 0.07 \text{ (syst)}.$$

In fact, measurement of the branching fraction of hadronic W decays to a final state containing a charm quark allows a direct measurement of the modulus of the CKM matrix element $V_{cs}$.

Neutrino-nucleon deep inelastic experiments (DIS) are complementary to charged lepton DIS experiments. In July 1998, Yang et al. [17] reported the extraction of structure functions by using the measurements of charged current neutrino and antineutrino nucleon interactions in the CCFR detector. The CKM matrix element $|V_{cs}|$ was extracted from a combined analysis of the structure function $xF_3$ and dimuon data. The resulting value of $|V_{cs}|$ is given by

$$|V_{cs}| = 1.05 \pm 0.10 \text{ (stat) } ^{+0.07}_{-0.11} \text{ (syst)}.$$

This determination is relatively free of theoretical assumptions. Systematic errors in these measurements are determined for each of the fit parameters by performing separate fits where experimental and model parameters are varied by their uncertainties. The measured value of the strange sea level parameter is in good agreement with the earlier value obtained from dimuon productions and yields the most precise measurement of $|V_{cs}|$.

There are two techniques for the precise measurement of $|V_{cb}|$. One of these involves measurements of the **inclusive** semileptonic branching fraction and lifetime to determine semileptonic rate of B meson that is proportional to $|V_{cb}|^2$. The constant of proportionality is obtained by theoretical quark-level calculations, enabling us to determine $|V_{cb}|$ with some uncertainties from hadronic effects. This method is based on the assumption that the inclusive sum is insensitive to the details of the various final states which contribute.

The second technique uses specific (**exclusive**) decays: either the mode $\bar{B} \to D^* l \bar{\nu}$ or the mode $\bar{B} \to D l \bar{\nu}$. *Inter alia,* the rate of decay for any one of these modes depends on strong interaction effects which are very difficult to quantify. Luke showed [18] that the first order correction vanishes for pseudoscalar to vector transitions. This makes theoretical analysis more precise for $\bar{B} \to D^* l \bar{\nu}$ than $\bar{B} \to D l \bar{\nu}$ for the determination of $|V_{cb}|$. Heavy Quark Effective Theory (HQET) [19-23] exploits the heavy quark symmetry and offers a rigorous framework for quantifying the hadronic effects with relatively small uncertainty [24-25]. The hadronic form factors in B semileptonic decays are known to be related to a single uniform factor, the Isgur-Wise function, by HQET that fixes its normalization at zero recoil of the charm meson. This property allows for an almost model-independent determination of the CKM matrix element $|V_{cb}|$ from the study of exclusive semileptonic B meson decays.

By 1996, all measurements of $|V_{cb}|$ based on semileptonic B meson decays had been performed from the differential decay rate of $\bar{B}^0 \to D^{*+} l^- \bar{\nu}_l$ [26-29]. The product of hadronic form factor and $|V_{cb}|$ is measured by extrapolation to point of zero recoil of the D meson and the value of $|V_{cb}|$ is then determined by using the theoretical prediction of the hadronic form factor. The theoretical uncertainty in this determination is of the order 3%. Although experimentally more difficult to measure, the semileptonic decay $\bar{B}^0 \rightarrow D^+l^- \bar{\nu}_l$ can also be used to determine $|V_{cb}|$. The corresponding hadronic form factor is predicted the same way and the theoretical uncertainty is of the same order.

The ALEPH Collaboration [30] reported the measurement of $|V_{cb}|$ in October 1966 by using both the exclusive decays. From approximately 3.9 million hadronic Z decays collected by the ALEPH detector at LEP, the Collaboration selected samples of two exclusive semileptonic decays, 579 $\bar{B}^0 \rightarrow D^{*+}l^- \bar{\nu}_l$ events and 261 $\bar{B}^0 \rightarrow D^+l^- \bar{\nu}_l$ events. The modulus of the CKM matrix element $V_{cb}$ is measured from the reconstructed differential decay rate of each sample. The value was extracted from the two samples using theoretical constraints on the slope and curvature of hadronic form factors and their renormalization at zero recoil. The value obtained was:

$$|V_{cb}| = (34.4 \pm 1.6 \text{ (stat)} \pm 2.3 \text{ (syst)} \pm 1.4 \text{ (th)}) \times 10^{-3}.$$

CLEO Collaboration [31], Adam et al., has determined $|V_{cb}|$ using a sample of $3 \times 10^6$ $B\bar{B}$ events in the CLEO detector at the Cornell Electron Strorage Ring. They have elaborated their early version and determined the yield of reconstructed $\bar{B}^0 \rightarrow D^{*+}l^- \bar{\nu}_l$ and $B^- \rightarrow D^{*0}l^- \bar{\nu}_l$ decays. They combined the product of $|V_{cb}|$ and the value of hadronic form factor at the kinematic end point at which the D* is at rest relative to the B, with the theoretical result for the letter obtained by lattice calculations [32] and determined

$$|V_{cb}| = 0.0469 \pm 0.0014 \text{ (stat)} \pm 0.0020 \text{ (syst)} \pm 0.0018 \text{ (theo)}.$$

This value of $|V_{cb}|$ is consistent with the previous values [30,33-35] but is somewhat higher. This value is also a bit higher than that obtained by using inclusive semileptonic B decays [36].

Bigi et al. [37] first pointed out that theoretical calculations including QCD corrections disagreed with the experimental results for inclusive semileptonic branching fraction of B decay. While the experimental value of branching fraction in various measurements has been less than 11% [38-40], theoretical expectations have been greater than 12%. The semileptonic branching fraction is combined with the decay lifetime to obtain the partial width that is used to extract the matrix element $|V_{cb}|$.

It is necessary for this measurement that primary decay leptons produced from $B \rightarrow X^+l^+\nu$ may be distinguished from secondary decay leptons produced mainly through decay $B \rightarrow \bar{D}X$, $\bar{D} \rightarrow Yl^- \bar{\nu}$). The Collaboration has used a dilepton method introduced by ARGUS [38] to minimize model dependence in this measurement. This approach requires a high momentum lepton, an electron or a muon, to identify a event and tag the flavour of one of them. The spectrum sdudy was performed using electrons only, as their experimental identification extends to much lower momenta than that of muons.

In August 2002, the Belle Collaboration, Abe et al. [35], reported a measurement of the electron spectrum from inclusive semi-leptonic B decay. A high-momentum lepton tag was used to separate the semi-leptonic B decay electrons from secondary decay electrons. The Collaboration obtained the fraction $\beta(B \rightarrow Xe^+\nu) =$ (10.90 ± 0.12 (stat) ± 0.49 (syst))% with minimal model dependence. From this measurement and world average of B meson lifetime, the value of the modulus of the matrix element $V_{cb}$ is derived as

$$|V_{cb}| = 0.0408 \pm 0.0010 \text{ (exp)} \pm 0.0025 \text{ (th)}.$$

The measurements of the magnitudes of the elements in the second row of the CKM matrix are not precise. The errors are so large that the sum of the squares of the magnitudes of the elements in the second row of the matrix does not leave the domain of the unitarity. For the third row, the desired data is not available.

We conclude that only more precise data can enable us to determine whether the unitarity of the CKM matrix is violated or not.

We may take this opportunity to mention the all important question: What is the source of CP violation? The question has yet remained unanswered. Gilman [41] has approached the problem in a phenomenological manner and has stated that it is the CKM matrix of the standard model that is the origin for this violation. He has asserted that the question to be answered by experiment and theory in the coming decade is: Are their CP-violating effects which do not arise from the CKM matrix and instead come from physics beyond the standard model? It appears that the matter has to be probed further and a convincing answer has yet to be sought.